\documentclass[10pt,conference]{IEEEtran}
\usepackage{cite}
\usepackage{graphicx}
\usepackage{amsmath,amssymb}
\usepackage{booktabs}
\usepackage{array}
\usepackage{xurl}
\usepackage[table]{xcolor}
\usepackage{hyperref}
\usepackage{tikz}
\usepackage{fontawesome5}
\usetikzlibrary{arrows.meta,positioning}
\usepackage{pgfplots}
\pgfplotsset{compat=1.18}
\definecolor{corpus}{HTML}{28666B}
\hypersetup{hidelinks}
\usepackage{balance}

\newcommand{\rqsummary}[2]{
  \par\medskip
  \begingroup
  \setlength{\fboxsep}{6pt}
  \noindent\colorbox{gray!15}{
    \parbox{\dimexpr\columnwidth-2\fboxsep\relax}{
      \textbf{#1 Summary.} #2}}
  \endgroup
  \par\medskip
}

\begin{document}

\title{\fontsize{20}{24}\selectfont Understanding Post-Quantum Cryptography Adoption: Practices, Challenges, and Evolution in Open-Source Software}

\author{
\IEEEauthorblockN{Souhardya Saha Dip$^{*}$,
Moshood A. Fakorede$^{*}$,
Krishna Upadhyay,
Umar Farooq}
\IEEEauthorblockA{Louisiana State University, Baton Rouge, Louisiana\\
\{ssahad1, mfakor1, kupadh4, ufarooq\}@lsu.edu}
}

\maketitle
\renewcommand{\thefootnote}{}\footnotetext{$^{*}$These authors contributed equally to this work.}\renewcommand{\thefootnote}{\arabic{footnote}}

\begin{abstract}
Post-quantum cryptography (PQC) standards establish algorithms for quantum-safe systems, but empirical evidence about how software projects implement and maintain PQC support remains limited.
We investigate this implementation and maintenance work by mining 114,361 commits and 5,713 issues across 179 open-source repositories. 
We use human annotation and validated LLM classification to identify PQC algorithms in commits, classify their adoption status and deployment mode, and identify technical challenges in issue discussions.
The commit classification for PQC algorithms reveals that 90\% of commits concern maintenance of existing support.
From the issue discussions, we construct an 11-category taxonomy of recurring technical challenges. 
Build, packaging, and dependency failures, along with functional defects, account for approximately half of categorized issues.
To examine how PQC support evolves, we use time-series analysis and first implementations of individual algorithms within repositories around NIST standardization milestones. 
After the 2024 finalization, standardized algorithm names appeared in more than 80\% of the PQC development changes analyzed, although earlier names such as Kyber continued to appear. 
Development activity increased over the observation period. Our milestone analysis, however, did not find a significant increase in first implementations. Migration plans should therefore account for sustained maintenance, prioritize build reliability and integration testing, and track both standardized and predecessor algorithm names. We make the dataset, taxonomy, annotations, and analysis pipeline publicly available.

\end{abstract}


\section{Introduction}
\label{sec:introduction}

Post-quantum cryptography (PQC), i.e., quantum-resistant cryptography, comprises algorithms designed to withstand attacks by classical and quantum computers~\cite{nist2016callforproposals}. A sufficiently large quantum computer would compromise widely used public-key algorithms that protect TLS, code signing, software updates, and VPNs~\cite{shor1994algorithms}. To prepare for this transition, the U.S. National Institute of Standards and Technology (NIST) conducted a multi-round standardization process~\cite{nist2016callforproposals} and finalized ML-KEM, ML-DSA, and SLH-DSA in August 2024~\cite{nist2024pqcstandardsannounce}, and selected HQC for standardization in March 2025~\cite{nist2025ir8545fourthround}. Migration must begin before cryptographically relevant quantum computers become available because adversaries can retain encrypted traffic for later decryption, while replacing cryptography across software ecosystems takes years~\cite{10.1109/MSP.2018.3761723,blancoromero2026practicalfeasibilityharvestnowdecryptlater}.

For software developers, adopting these algorithms involves implementation, integration, and continued maintenance. Projects need to make PQC code work with existing protocols, dependencies, build systems, and platforms. They may support PQC algorithms independently or combine them with classical algorithms in hybrid modes. Developers must also track changes between candidate algorithms and their standardized successors. For example, ML-KEM is based on CRYSTALS-Kyber~\cite{nist2024pqcstandardsannounce}. Understanding adoption therefore requires evidence about how projects introduce support, maintain it, and respond to technical problems over time.

Prior empirical studies measure PQC readiness, support, and deployment through binary analysis of applications~\cite{strauss2025assessingenhancingquantumreadiness}, network measurements of negotiated post-quantum protocols~\cite{loizou2026measuringpostquantumtlsdeployment,wickramasinghe2026mindgappolicyvs}, and static analysis of PQC API usage in Python, Java, and Go repositories~\cite{mehl2026pqcreadiness}. Industry reports from Meta, Apple, AWS, Chromium, and Microsoft describe performance, interoperability, dependency, and rollout constraints in their systems~\cite{meta2024pqrtls,meta2026pqcmigrationframework,apple2024pq3,aws2024pqcmigrationplan,chromium2026pqroadmap,microsoft2025dotnetpqc}. Migration reviews organize reported transition processes and challenges~\cite{nather2024migratingsoftwaresystems}.

Our contribution is a cross-project characterization of PQC implementation, maintenance, recurring technical challenges, and evolution. We mine 114,361 commits and 5,713 issues across 179 public GitHub repositories~(Fig.~\ref{fig:study-overview}).
The corpus includes PQC-focused projects and general-purpose systems such as OpenSSL~\cite{openssl2026} and Bouncy Castle~\cite{bouncycastle2026}. Commit patches, changed paths, and linked merged pull requests distinguish new implementations from changes to existing support. Issue text, comments, and available resolving pull requests document technical problems. Human annotation are used to establish reference labels and the technical-challenge taxonomy. LLM classifiers validated against held-out human labels classify the remaining selected records.

We classify each affected PQC algorithm separately within a commit to characterize observed source-level support. Adoption-status labels distinguish implementation and maintenance from discussion, planning, deprecation or removal, and rejection or reversion. Deployment-mode labels record whether the source evidence indicates pure PQC, a hybrid construction combining PQC and classical algorithms, or both configurations. We count implementation and maintenance as adoption evidence. Temporal analysis separates continued development from the first implementation of each algorithm within a repository around NIST milestones.

We address three research questions about adding PQC support, maintaining it, and tracking its development: 

\begin{figure*}[!t]
\centering
\includegraphics[width=0.85\textwidth]{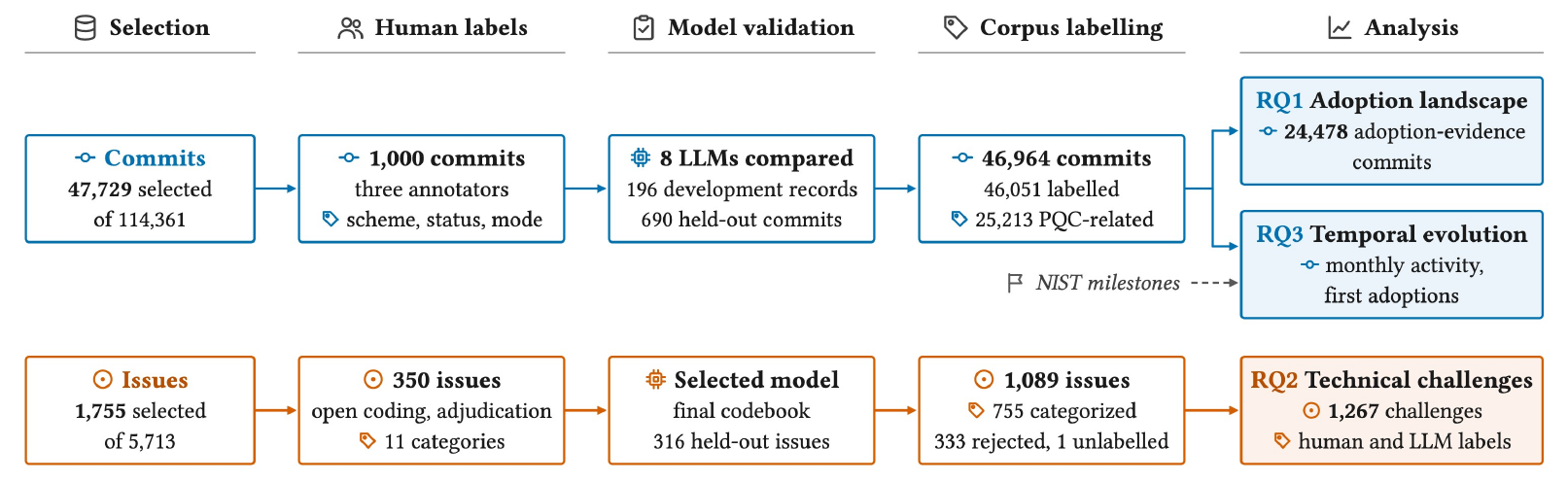}
\vspace{-1em}
\caption{Study design: Commits (top) and issues (bottom) follow the same validated LLM classification protocol, from human reference labels through model validation to corpus labeling. RQ1 and RQ3 analyze commits, and RQ2 analyzes issues (record counts reported at each step).}
\label{fig:study-overview}
\vspace{-1em}
\end{figure*}

\begin{itemize} 
\item Evidence about supported algorithms and their configurations helps projects assess PQC integration options.  \textbf{RQ1.} Which PQC algorithms appear in open-source development, and what are their adoption statuses and deployment modes? 

\item Problems that recur while projects establish and maintain this support can inform maintenance and testing priorities. \textbf{RQ2.} What recurring technical challenges do developers report when implementing, integrating, and migrating to PQC? 

\item Tracking development activity separately from each algorithm's first working implementation within a repository helps assess how standardization aligns with the expansion of support. \textbf{RQ3.} How do PQC development activity and first algorithm implementations within repositories change around NIST standardization milestones?
\end{itemize}

\noindent\textbf{Key Findings.}
Across the studied repositories, developers continue to modify PQC support after its initial implementation.
\textbf{Maintenance accounts for 90.6\%} of the algorithm-specific adoption-status labels, compared with 3.8\% for new implementations. 
This activity spans pure-PQC and hybrid configurations, with ML-KEM and ML-DSA supported by the most repositories. In issue discussions, developers most often report \textbf{build, packaging, and dependency failures and functional defects}. Together, these categories account for 49.8\% of the 1,267 issues categorized in our 11-category taxonomy. 
They recur across the most projects and retain the largest shares when macro-averaged across repositories. Integration challenges occur less often but have a longer estimated median time to recorded closure than build failures in an unadjusted comparison. 
After NIST finalized the first standards in 2024, standardized algorithm names dominate recorded activity, while predecessor names such as Kyber persist. 
The baseline analysis associates the 2024 finalization with increased implementation and maintenance activity, although the estimate is sensitive to timing, exposure, and autocorrelation treatment. 
When we count each algorithm's first working implementation within a repository, we find \textbf{no statistically distinguishable increase in first-implementation rates} at the milestones. 
The maintenance results and challenge taxonomy support migration plans that budget for maintenance after initial implementation and prioritize build reliability and integration testing. Since predecessor names persist, cryptographic inventories should record both predecessor and standardized algorithm names while preserving their versions.

To support further research, we provide analyzed records, codebooks, human annotations, classifier prompts and outputs, and supporting code in the replication package~\cite{pqcAdoptionReplicationPackage}.

\section{Study Design}
\label{sec:methodology}

Our study combines repository mining, human annotation, and validated LLM
classification to study PQC adoption and technical challenges
(Figure~\ref{fig:study-overview}).

\subsection{Data Collection}
\label{sec:methodology-discovery}
\label{sec:methodology-tiering}
\label{sec:methodology-mining}

We collected commits, pull requests (PRs), and issues from 179 public GitHub
repositories.

\noindent\textbf{Repository selection.} We manually selected 50 repositories from
\emph{Awesome Quantum Software}, \emph{Awesome PQC Resources},
\emph{Awesome Post-Quantum}, and projects recommended by PQClean~\cite{pqclean2026}.
The curated set covered general-purpose projects such as
OpenSSL~\cite{openssl2026} and Bouncy Castle~\cite{bouncycastle2026},
including projects without PQC in their descriptions.

Three GitHub discovery channels expanded this set in September 2026.
Topic searches used \texttt{topic:} for 25 PQC topics, split by creation year
over 2016--2026 to stay below GitHub's 1,000-result limit. Metadata searches
used \texttt{in:name,description,topics} with a 95-term PQC vocabulary,
joining up to six unambiguous terms, such as ML-KEM, FrodoKEM, and FIPS~203,
with OR. Ambiguous names such as Falcon required a context phrase such as
post-quantum. The third channel retrieved repositories owned by PQC
organizations, including \texttt{open-quantum-safe}, \texttt{pq-code-package},
and \texttt{mupq}.

Search-discovered repositories required at least 30 stars and PQC-related metadata;
forks and archived or disabled repositories were excluded. These filters
did not apply to the curated set. Deduplication and screening yielded
129 additional repositories after 18 manual exclusions for reasons such as
misleading names, absent PQC content, or non-software material.

\noindent\textbf{Collection scope.} For each repository, we computed the share of
PRs and issues matching any of 90 PQC terms in their titles, bodies, or comments.
This vocabulary omitted five ambiguous discovery terms; records matching
multiple terms counted once.
The collection threshold was 0.148, the lowest matching share among 24
PQC-focused reference repositories. All 17 general-purpose reference
repositories fell below it.

We collected all PRs, issues, and commits from the 81 repositories at or
above this threshold (Corpus~A). For the 87 below it (Corpus~B), we collected
PRs and issues only when their title or body matched a keyword, and commits
with a message match or a link to a selected PR. The remaining 11
repositories had no PRs or issues and contributed only Git commit histories (Corpus~C).

Frozen mirrors were used to preserve Git histories for RQ1 and RQ3. For collection, we used
GitHub's REST API for PR and issue records and its GraphQL API for links
between PRs, commits, and closed issues. All comment and review pages for
selected threads were retrieved without further keyword filtering.
Collection ran over September 21--24, 2026, retaining author types and
original timestamps from both new records and earlier PR and issue inventories.

The candidate corpus contained 114{,}361 repository--commit pairs,
11{,}919 PRs, and 5{,}713 issues (Section~\ref{sec:results}). The artifact
documents queries, selection decisions, and collection history.

\subsection{Commit Selection and Filtering}
\label{sec:methodology-commit-filtering}

Commits were eligible if reachable from the default branch or linked to a
merged PR, leaving 73{,}076 repository--commit pairs. We removed 2{,}175 rebased
copies by matching patches and commit metadata. Patch identifiers, following
German et al.\ \cite{german2016continuously}, identified another 2{,}443 merge
commits that repeated a patch and 1{,}587 PR commits preserved by a squash merge
on the default branch. Removing these left 66{,}871 commits. Another
3{,}067 merges were excluded because their branch tips were already
represented in the corpus.

In the OQS OpenSSH fork, 11{,}375 inherited commits had no PQC term in the
message, linked PR or issue text, file paths, or changed lines and were
excluded. The fork retained 2{,}256 commits mentioning terms such as Kyber,
ML-KEM, liboqs, or sntrup. Applying the same keyword
check to repositories whose names, descriptions, and topics did not indicate
PQC removed another 1{,}934 commits.

Rule-based filtering excluded 2{,}766 of the remaining 50{,}495 commits as
non-PQC. These were empty commits or commits with no PQC term in the diff
that were bot-authored or changed only CI configuration, lock files, or
repository metadata such as licence files. The final classification corpus contained 47{,}729 commits
from 163 repositories.

\subsection{Analysis Overview}
\label{sec:methodology-analysis-overview}

Human annotations were used to establish reference labels for classifying the commits and constructing the technical-challenge
taxonomy. Validated models then classified the remaining records. RQ1 examined
scheme adoption, RQ2 recurring challenges and time to recorded issue closure,
and RQ3 temporal trends in adoption and development activity.

\subsection{LLM-Based Validation}
\label{sec:methodology-llm-validation}

Prior LLM-based evaluation only addressed PQC code-fragment migration
\cite{debonrostro2026empiricalevaluationlargelanguage}. For commit and issue
classification, we evaluated
eight proprietary and open-weight models from top AI labs released
between March 17 and October 7, 2026, comprising Gemini 3.8 Flash
\cite{google2026gemini38flash}, Claude Haiku 5.5
\cite{anthropic2026claudehaiku55}, GPT-5.4 mini
\cite{openai2026gpt54mini}, Muse Spark 1.3
\cite{meta2026musespark13}, GLM-5.3 \cite{zai2026glm53}, Kimi K2.7 Code
\cite{moonshot2026kimik27code}, Qwen3.8-Flash
\cite{alibaba2026qwen38flash}, and Mistral Medium 3.5
\cite{mistral2026medium35}.

Comparisons were ran through OpenRouter~\cite{openrouter2026}
between October 5 and 11, 2026, with medium reasoning effort where supported
and high effort for GLM-5.3. Output limits were 8{,}192 tokens,
increased to 32{,}768 for GLM-5.3 and Kimi K2.7 Code. The artifact provides
API identifiers and inference settings following reporting guidance
\cite{baltes2026guidelinesempiricalstudies}.

Within each comparison, models received identical codebooks, prompts,
evidence, and output schemas. Model selection and prompt refinement used
development data. The model, prompt, and evaluation criteria were then frozen
before a single held-out evaluation to limit selection bias
\cite{cawley2010overfittingmodelselection}. Evaluation covered precision,
recall, macro-F1, and per-class F1, with class counts and bootstrap confidence
intervals, following recommendations for
LLM-supported software-engineering research
\cite{ahmed2025llmsreplacemanualannotation,
bano2024largelanguagemodelsqualitative,
baltes2026guidelinesempiricalstudies}. Corpus classification required the
model to meet the acceptance criteria in
Sections~\ref{sec:rq1} and~\ref{sec:rq2}.

\section{Empirical Study}
\label{sec:results}

In this section, we examine scheme adoption (RQ1), recurring technical challenges (RQ2),
and development activity and first adoption around NIST milestones (RQ3).

\subsection{RQ1 Adoption Landscape}
\label{sec:rq1}
\label{sec:results-rq1}
\label{sec:methodology-classification}

\subsubsection{Approach}

\textbf{Unit and evidence.} Each record is a repository--commit pair from
the filtered corpus (Section~\ref{sec:methodology-commit-filtering}), with
patches, changed paths, commit messages, linked merged PRs, and issues closed
by those PRs. The prompt prioritized patches over prose and prohibited reliance
on repository reputation or external knowledge.

The labels cover commit-level PQC relevance and each changed scheme's adoption
status, deployment mode, classical hybrid partner, and confidence.
Adoption statuses were \emph{implemented}, \emph{maintenance},
\emph{planned}, \emph{discussed}, \emph{deprecated or removed},
\emph{rejected or reverted}, and \emph{unclear}. Only the first two counted
as adoption evidence. New working support counted as \emph{implemented};
benchmarks, test vectors, regenerated files, and partial changes as
\emph{maintenance}; scanners, catalogues, and comparison tables as
\emph{discussed}. Deployment modes were \emph{hybrid}, \emph{pure},
\emph{both}, \emph{not applicable}, and \emph{unclear}. Hybrid labels required
classical and PQC components in the same mechanism. Predecessor and
standardized names remained distinct, including Kyber and ML-KEM, and
Dilithium and ML-DSA.

\noindent\textbf{Human annotation.} We drew a repository-aware, disproportionate
stratified sample of 1{,}000 commits following repository-sampling
and diversity guidance
\cite{gorostidi2024creationrepresentativesamplessoftware,10.1145/2491411.2491415}.
Sampling balanced scheme strata, merged-PR and default-branch evidence, and
creation periods, with at most 20 records per repository. Rare schemes were
oversampled, so unweighted sample proportions did not estimate corpus prevalence.
The pilot, development, and held-out sets contained 100, 200, and 700 records.
Three authors independently labelled the pilot with the same
codebook. All had at least seven years of software-engineering experience;
one had three years of quantum-computing experience and two had one year each.
Relevance used Fleiss' and pairwise Cohen's $\kappa$ and
Krippendorff's $\alpha$; scheme sets used exact match and mean Jaccard overlap.
Status and mode used Fleiss' $\kappa$ on 103 commit--scheme pairs named by
all three annotators (60 pilot commits).

\noindent\textbf{Model selection.} The eight models in
Section~\ref{sec:methodology-llm-validation} received the same initial prompt,
codebook, and evidence, including repository descriptions but excluding record
identifiers, corpus assignments, sampling strata, and keyword matches.
The prompt required one status and mode per scheme. Evaluation covered 196
development records after excluding three oversized patches and one encoding
failure; invalid responses and over-context requests counted as errors.

Relevance was evaluated using Cohen’s $\kappa$ and F1 for the PQC-related class. Scheme identification used micro-F1 over commit–scheme pairs. Status and mode $\kappa$
covered schemes present in both model and human labels, with equal
commit weights. We estimated 95\%
confidence intervals and paired model differences from 1{,}000 record-level
bootstrap resamples with a fixed seed, applying Holm correction to model
comparisons. Sensitivity to scheme-stratum composition used record weights
$w_i = N_{h(i)}/n_{h(i)}$,
where $h(i)$ is record $i$'s primary scheme stratum, $N_h$ counts commits in
that stratum of the original 73{,}076-commit sampling frame before subsequent
filtering, and $n_h$ counts scored development records in stratum $h$.
The weights omit repository caps and coverage constraints, so they support
sensitivity analysis rather than exact inverse-probability estimates.
Weighted relevance, status, and mode $\kappa$, adoption-evidence Jaccard
overlap, and adoption-evidence micro-F1 used within-stratum bootstrap
resampling. Together with scheme micro-F1, these formed six ranking metrics
$M$. The mean rank, averaging ties within each metric, was
\begin{equation}
\label{eq:mean-rank}
\bar{r}_m = \frac{1}{|M|}\sum_{k \in M} \mathrm{rank}_k(m).
\end{equation}
The three models with the lowest $\bar{r}_m$ advanced.

\noindent\textbf{Prompt revision and alternative annotator test.} Revisions based on
development-set disagreements were used to clarify patch-based decisions, implementation
versus maintenance or discussion, and pure deployment as the default for standalone
implementations. The revised prompt required labels for every scheme affected by the commit. Build and packaging fixes to existing support were labelled maintenance.

Using the revised prompt, the pilot alternative annotator
test~\cite{calderon2025alternativeannotatortest} compared model $f$ and annotator
$j$ against the remaining annotators $K_j$ for each commit $i$,

\begin{equation}
\label{eq:alt-alignment}
\begin{aligned}
S^{f}_{ij} &= \frac{1}{|K_j|}\sum_{k \in K_j} \mathrm{sim}\big(f(x_i), h_k(x_i)\big),\\
S^{h}_{ij} &= \frac{1}{|K_j|}\sum_{k \in K_j} \mathrm{sim}\big(h_j(x_i), h_k(x_i)\big),
\end{aligned}
\end{equation}
where $\mathrm{sim}$ was exact match for relevance, Jaccard overlap for scheme
and adoption-evidence sets, and the agreement share for status and mode over
schemes named by all three annotators. For status and mode, model omissions
counted as disagreements; commits without such schemes were excluded.

Model wins were
$W^{f}_{ij} = \mathbb{1}[S^{f}_{ij} \geq S^{h}_{ij}]$ and annotator wins were
$W^{h}_{ij} = \mathbb{1}[S^{h}_{ij} \geq S^{f}_{ij}]$, counting ties for both.
Their means over commits were $\rho^{f}_{j}$ and $\rho^{h}_{j}$.
We tested $H_{0j}\colon \rho^{h}_{j} - \rho^{f}_{j} \geq \epsilon$ with
one-sided paired $t$-tests on $W^{h}_{ij} - W^{f}_{ij}$, applying the
Benjamini--Yekutieli correction~\cite{benjamini2001fdrdependency} to the three
annotator comparisons, using $q = 0.05$ divided across the three candidate
models by Bonferroni correction~\cite{dunn1961multiplecomparisons}.
Using $q = 0.05$ per model gave the same outcomes.
A field passed when the share $\omega$ of rejected annotator-specific null
hypotheses reached 0.5. Mean advantage probability was
$\bar{\rho} = \frac{1}{3}\sum_j \rho^{f}_{j}$. The pre-specified skilled-annotator
margin was $\epsilon = 0.15$, with sensitivity checks at 0.10 and 0.20.

\noindent\textbf{Corpus classification.} Excluding human-annotated records from the
filtered corpus (Section~\ref{sec:methodology-commit-filtering}) left
46{,}964 commits in 147 repositories for automated classification.
The passing model with the strongest agreement classified these records in
95 OpenRouter Batch API batches of at most 1{,}000 commits through Google Vertex.

\begin{figure*}[!t]
\begin{minipage}[t]{\columnwidth}
\centering
\includegraphics[width=\linewidth]{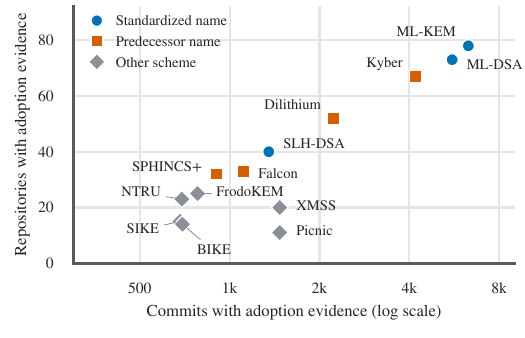}
\caption{Adoption breadth and commit activity for the 13 schemes with the most
adoption-evidence commits (\emph{implemented} or \emph{maintenance}).}
\label{fig:corpus-scheme-patterns}
\end{minipage}\hfill
\begin{minipage}[t]{\columnwidth}
\centering
\includegraphics[width=\linewidth]{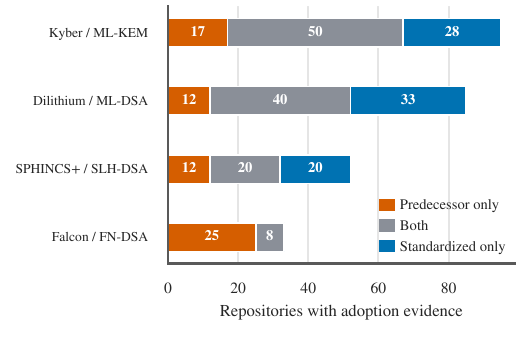}
\caption{Predecessor--standard coexistence. Repository counts by adoption of
predecessor names, standardized names, or both.}
\label{fig:scheme-name-coexistence}
\end{minipage}
\end{figure*}

\subsubsection{Result}

\textbf{Human agreement.} Fleiss' $\kappa$ indicated almost perfect agreement
for PQC relevance (0.819) and adoption status (0.900), and substantial agreement
for deployment mode (0.725)~\cite{landis1977measurement}
(Table~\ref{tab:pilot-agreement}). Scheme-set agreement was also high,
with Jaccard-based Krippendorff's $\alpha$ of 0.841.
The artifact reports pairwise and per-scheme diagnostics.

\begin{table}[!htb]
\centering
\caption{Human agreement on 100 pilot commits. Pairwise values are mean raw
agreement; scheme sets use exact matches and Jaccard-based $\alpha$.}
\label{tab:pilot-agreement}
\footnotesize
\begin{tabular}{@{}lcccc@{}}
\toprule
Field & Fleiss $\kappa$ & $\alpha$ & Pairwise agr. & Unanimous \\
\midrule
PQC relevance & 0.819 & 0.820 & 92.0\% & 88.0\% \\
Scheme set & n/a & 0.841 & 80.3\% & 73.0\% \\
Adoption status & 0.900 & 0.900 & 94.8\% & 92.2\% \\
Deployment mode & 0.725 & 0.726 & 90.0\% & 85.4\% \\
\bottomrule
\end{tabular}
\end{table}

\noindent\textbf{Model selection.} Gemini 3.8 Flash had the highest status and mode
$\kappa$ (Table~\ref{tab:model-selection}), with 95\% bootstrap intervals
of 0.717--0.919 and 0.648--0.861, respectively.
After Holm correction, Gemini exceeded Mistral, Kimi, Qwen, and GLM on status
$\kappa$, and Claude, Muse, Mistral, Kimi, and Qwen on mode $\kappa$.
Its relevance advantage was significant against Mistral and Qwen, which
failed on 24 and 23 of the 196 records, mainly after exhausting output limits.
Gemini and Muse tied at mean rank 2.17, followed by GPT-5.4 mini at 2.33
(Eq.~\eqref{eq:mean-rank}); these three advanced. The remaining models
scored at least 4.33.

\begin{table}[!htb]
\centering
\caption{Model selection on 196 development records using the initial prompt.
Cohen's $\kappa$ is record-weighted; failures count as errors.
Column bests are bold.}
\label{tab:model-selection}
\footnotesize
\setlength{\tabcolsep}{3pt}
\begin{tabular}{@{}lcccc@{}}
\toprule
Model & PQC F1 & Scheme F1 & Status $\kappa$ & Mode $\kappa$ \\
\midrule
Gemini 3.8 Flash & 0.936 & 0.860 & \textbf{0.832} & \textbf{0.762} \\
Muse Spark 1.3 & \textbf{0.940} & 0.870 & 0.766 & 0.547 \\
GPT-5.4 mini & 0.924 & \textbf{0.875} & 0.713 & 0.667 \\
Claude Haiku 5.5 & 0.915 & 0.734 & 0.703 & 0.591 \\
GLM-5.3 & 0.935 & 0.862 & 0.666 & 0.700 \\
Kimi K2.7 Code & 0.908 & 0.829 & 0.624 & 0.488 \\
Qwen3.8-Flash & 0.856 & 0.734 & 0.557 & 0.511 \\
Mistral Medium 3.5 & 0.841 & 0.676 & 0.498 & 0.506 \\
\bottomrule
\end{tabular}
\end{table}

\noindent\textbf{Prompt revision and alternative annotator test.} The prompt was
revised using development-set disagreements, then evaluated on 100 pilot
commits. Gemini's mean Cohen's $\kappa$ across comparisons with the three
human annotators was 0.830 for relevance, 0.828
for status, and 0.787 for mode. After Holm correction, Gemini exceeded
GPT-5.4 mini on status and mode $\kappa$, scheme micro-F1, and adoption-evidence
Jaccard overlap. Gemini and Muse did not differ significantly on any
unweighted metric. Full pilot comparison results are omitted here for space
and provided in the supplementary material.

In the alternative annotator test (Table~\ref{tab:alt-test}), Gemini and Muse
passed all five fields at $\epsilon = 0.15$; GPT-5.4 mini failed only status and mode.
Results were unchanged at $\epsilon = 0.20$.
At $\epsilon = 0.10$, Gemini and Muse passed relevance, schemes, and adoption
evidence; GPT-5.4 mini passed only relevance. We selected Gemini 3.8 Flash
for its highest $\bar{\rho}$ on status, mode, and adoption evidence, and higher
stratum-weighted mode $\kappa$ than Muse, 0.870 versus
0.766 (Holm-adjusted $p = 0.016$).

\begin{table}[!t]
\centering
\caption{Alternative annotator test using the revised prompt on the pilot
($\epsilon = 0.15$). Cells report mean
advantage probability $\bar{\rho}$ and winning rate $\omega$ (pass at
$\omega \geq 0.5$). Row-best $\bar{\rho}$ values are bold.}
\label{tab:alt-test}
\footnotesize
\setlength{\tabcolsep}{3pt}
\begin{tabular}{@{}lccc@{}}
\toprule
Field & Gemini 3.8 Flash & Muse Spark 1.3 & GPT-5.4 mini \\
\midrule
PQC relevance & \textbf{0.980 / 1.00} & \textbf{0.980 / 1.00} & 0.960 / 1.00 \\
Scheme set & \textbf{0.970 / 1.00} & 0.953 / 1.00 & 0.930 / 1.00 \\
Adoption status & \textbf{0.950 / 1.00} & 0.933 / 1.00 & 0.833 / 0.00 \\
Deployment mode & \textbf{0.950 / 1.00} & 0.939 / 1.00 & 0.889 / 0.00 \\
Adoption evidence & \textbf{0.970 / 1.00} & 0.940 / 1.00 & 0.910 / 1.00 \\
\bottomrule
\end{tabular}
\end{table}

\noindent\textbf{Held-out evaluation.} A single evaluation of the frozen classifier
covered 690 of the 700 human-labelled held-out records; 10 oversized or
non-UTF-8 commits were never submitted. Failed responses counted as errors.
Gemini reached F1 0.93 for PQC-related commits (precision 0.97, recall 0.89)
and 0.85 for non-PQC commits, giving macro-F1 0.89. Scheme micro-F1 was 0.84,
with precision 0.93 and recall 0.77. Status and mode macro-F1 were 0.70 and
0.68. At the commit--scheme level, \emph{rejected or reverted} had only
2 examples; \emph{unclear} deployment had 46 and recall 0.04.
Recall for \emph{deprecated or removed} was 0.58, while frequent classes
reached F1 of 0.80--0.89.

\noindent\textbf{Corpus classification.} Gemini labelled 46{,}051 of 46{,}964 commits
(98.1\%) for \$422.96. The 913 unlabelled records comprised 363 over-context
requests, 245 invalid answers (200 at the output limit), 218 oversized patches,
82 non-UTF-8 patches, and 5 reviewed failures. The classifier labelled
25{,}213 commits (54.8\%) as PQC-related and 20{,}838 as non-PQC. PQC-related commits
occurred in 144 of 147 repositories; the model assigned high confidence to
96.0\% of them.

\subsubsection{Observation}

Scheme implementation or maintenance accounted for 24{,}478 (97.1\%) of the
PQC-related commits, covering 135 repositories, 107 schemes, and 834 repository--scheme
adoption pairs. Lattice-based standards had the broadest adoption
(Figure~\ref{fig:corpus-scheme-patterns}). ML-KEM appeared in 78 repositories,
ML-DSA in 73, Kyber in 67, and Dilithium in 52.

In all four renamed lineages, at least half of repositories with standardized
names also had predecessor names
(Figure~\ref{fig:scheme-name-coexistence}). Kyber or ML-KEM evidence appeared
in 95 repositories, including 50 with both and 17 with Kyber only. Dilithium and ML-DSA
appeared in 85 repositories, including 40 with both; SPHINCS+ and SLH-DSA
appeared in 52, including 20 with both. All 8 repositories with FN-DSA also
contained Falcon, which appeared in 33.

Picnic had 1{,}471 adoption-evidence commits in 11 repositories, whereas
1{,}353 SLH-DSA commits spanned 40 (Figure~\ref{fig:corpus-scheme-patterns}).

Maintenance accounted for 90.6\% of the 34{,}937 scheme labels, followed by
\emph{implemented} (3.8\%), \emph{discussed} (3.5\%), \emph{deprecated or
removed} (1.6\%), and \emph{rejected or reverted} and \emph{planned}
(0.2\% each). RQ3 used the 1{,}331 \emph{implemented} labels to identify
first-adoption events.

Deployment modes across the 32{,}994 adoption labels were 90.1\% pure,
4.6\% hybrid, 4.3\% both, and 1.0\% unclear. The 2{,}932 hybrid or both labels spanned 51 repositories.
X25519 was the leading classical partner in 908 commits, followed by P-256
in 416, P-384 in 352, Ed25519 in 348, and P-521 in 190. Hybrid key exchange
centered on Curve25519 and NIST curves; hybrid signatures centered on Ed25519.

\rqsummary{RQ1}{ML-KEM (78 repositories) and ML-DSA (73) had the broadest
adoption, while predecessor schemes persisted in all four renamed lineages.
Pure-PQC deployment dominated (90.1\% of adoption labels); hybrid or mixed
modes accounted for 8.9\% across 51 repositories. Maintenance comprised
90.6\% of scheme labels.}

\subsection{RQ2 Recurring Technical Challenges}
\label{sec:rq2}
\label{sec:results-rq2}
\label{sec:methodology-taxonomy}

\subsubsection{Approach}

\textbf{Unit and eligibility.} Each record comprised an issue's metadata,
opening text, chronological comments, and, where available, the closing PR's
text, comments, and commit messages. Coders and the classifier received
the same evidence; linked PRs provided context and were not counted separately.

Eligibility required a PQC-related technical problem or obstacle, excluding
feature requests, usage questions without a reported problem, announcements,
routine maintenance, and incidental keyword matches. Three researchers
independently screened the same 100 pilot
issues; each of the 200 development and 300 screening held-out issues received
one human label. Agreement estimates used the pilot; the frozen screening
classifier was evaluated on all 300 held-out issues. Eligibility was checked
again during taxonomy coding.

\noindent\textbf{Taxonomy construction.} Two researchers independently open-coded
issues that had passed initial screening~\cite{seaman1999qualitativemethods}.
Coding proceeded in four rounds of 50, 100, 100, and 100 issues. Sampling used a
fixed seed and previously uncoded issues, balancing repositories, creation
years, schemes, ecosystems, and project roles. Coders assessed eligibility,
described each eligible problem, and assigned an open code. From round 2,
they also selected a primary category and optional sub-category for the
motivating problem, or proposed a new category. The coders grouped related codes
into categories through inductive (bottom-up) thematic
synthesis~\cite{cruzes2011thematicsynthesis}.

Pre-adjudication agreement was measured with Cohen's $\kappa$. A third researcher chaired
meetings to resolve disagreements and revise definitions.
Discovery stopped after two consecutive rounds added no primary category,
adapting thematic-saturation guidance
\cite{guest2020simple}. Taxonomy construction took approximately 18 hours.

\begin{table*}[!t]
\centering
\caption{Taxonomy of recurring PQC technical challenges across 1{,}267 categorized issues.}
\label{tab:rq2-taxonomy}
\footnotesize
\setlength{\tabcolsep}{4pt}
\renewcommand{\arraystretch}{1.08}
\begin{tabular}{@{}>{\raggedright\arraybackslash}p{3.6cm} >{\raggedright\arraybackslash}p{\dimexpr\textwidth-11.6cm-4\tabcolsep\relax} >{\raggedright\arraybackslash}p{8.0cm}@{}}
\toprule
\rowcolor{gray!12}
Category (code), issues (\%) & Operational definition & Examples (condensed; source issue as repo\#issue) \\
\midrule
\rowcolor{gray!7}
\textbf{Build, packaging, and dependency failures} (C01), 391 (30.9\%) & PQC code cannot be compiled, linked, installed, or resolved as a dependency, or the installed package is incomplete & Dropping old Kyber breaks the s2n-tls build.~{\scriptsize(\href{https://github.com/aws/s2n-tls/issues/5699}{s2n-tls\#5699})} \newline Can't use slh-dsa and ml-dsa due to pre releases.~{\scriptsize(\href{https://github.com/RustCrypto/signatures/issues/930}{signatures\#930})} \\
\rowcolor{gray!7}
\multicolumn{3}{@{}p{\textwidth}@{}}{\scriptsize\textit{Subcategories.} Platform builds; configuration builds; dependency and packaging; source defects} \\
\addlinespace[2pt]
\textbf{Incorrect results and functional defects} (C04), 240 (18.9\%) & The implementation gives wrong results, hangs, or misses a case it should support, with no dispute about the standard & ML-KEM public-key equality check is wrong.~{\scriptsize(\href{https://github.com/openssl/openssl/issues/28563}{openssl\#28563})} \newline In-place signing failure wipes the input message.~{\scriptsize(\href{https://github.com/pq-code-package/mldsa-native/issues/1217}{mldsa-native\#1217})} \\
\addlinespace[2pt]
\rowcolor{gray!7}
\textbf{Protocol and ecosystem integration} (C07), 124 (9.8\%) & A scheme cannot be used in an existing protocol, product, or deployment because identifiers, library versions, or integration pieces are missing or incompatible & BCJSSE on Java 25 disables ML-KEM groups.~{\scriptsize(\href{https://github.com/bcgit/bc-java/issues/2252}{bc-java\#2252})} \newline Go's x509 cannot verify PQC-signed certificates.~{\scriptsize(\href{https://github.com/golang/go/issues/74024}{go\#74024})} \\
\addlinespace[2pt]
\textbf{Platform-specific runtime faults} (C02), 112 (8.8\%) & Code builds but crashes or gives wrong results on one platform, compiler, runtime, or memory configuration & Stack overflow generating a key pair on Windows.~{\scriptsize(\href{https://github.com/open-quantum-safe/liboqs-rust/issues/134}{liboqs-rust\#134})} \newline Assembly code fails on aarch64 with BTI enabled.~{\scriptsize(\href{https://github.com/smuellerDD/leancrypto/issues/38}{leancrypto\#38})} \\
\multicolumn{3}{@{}p{\textwidth}@{}}{\scriptsize\textit{Subcategories.} Memory and stack limits; compiler and toolchain faults; ABI and alignment faults} \\
\addlinespace[2pt]
\rowcolor{gray!7}
\textbf{Security weaknesses} (C05), 102 (8.1\%) & The implementation or interface leaks secrets, accepts invalid input, mishandles keys, or makes unsafe use easy & Variable-time division in Kyber decapsulation.~{\scriptsize(\href{https://github.com/mupq/pqm4/issues/319}{pqm4\#319})} \newline AVX2 verifier accepts some invalid signatures.~{\scriptsize(\href{https://github.com/pq-crystals/dilithium/issues/113}{dilithium\#113})} \\
\rowcolor{gray!7}
\multicolumn{3}{@{}p{\textwidth}@{}}{\scriptsize\textit{Subcategories.} Side channels; key handling; input validation; misuse risk} \\
\addlinespace[2pt]
\textbf{Specification conformance and interoperability} (C06), 102 (8.1\%) & Behaviour, results, or encodings differ from the standard, another scheme version, or the reference implementation & Encapsulate's output order differs from FIPS 203.~{\scriptsize(\href{https://github.com/golang/go/issues/70950}{go\#70950})} \newline ML-DSA key format changed between IETF drafts.~{\scriptsize(\href{https://github.com/randombit/botan/issues/5002}{botan\#5002})} \\
\multicolumn{3}{@{}p{\textwidth}@{}}{\scriptsize\textit{Subcategories.} Draft versions; specification deviation; reference tracking and test vectors (KAT mismatches); representation differences; specification text errors} \\
\addlinespace[2pt]
\rowcolor{gray!7}
\textbf{Test infrastructure and test vectors} (C10), 72 (5.7\%) & Tests cannot run or cannot be trusted because data, fixtures, or infrastructure are missing or broken & Tests report `All tests passed!' despite failures.~{\scriptsize(\href{https://github.com/SQISign/the-sqisign/issues/10}{the-sqisign\#10})} \newline Wycheproof checks are skipped under python -O.~{\scriptsize(\href{https://github.com/pq-code-package/mldsa-native/issues/1218}{mldsa-native\#1218})} \\
\addlinespace[2pt]
\textbf{Performance and efficiency} (C11), 59 (4.7\%) & A scheme, implementation, or test is too slow, or a benchmark misreports speed or memory & ML-DSA-44 signing p99 latency tripled.~{\scriptsize(\href{https://github.com/paulmillr/noble-post-quantum/issues/40}{noble-post-quantum\#40})} \newline FrodoKEM is slow: AES runs one block at a time.~{\scriptsize(\href{https://github.com/randombit/botan/issues/4173}{botan\#4173})} \\
\addlinespace[2pt]
\rowcolor{gray!7}
\textbf{Formal verification and proof obstacles} (C09), 42 (3.3\%) & A formal proof is blocked, or a formal statement or model is wrong & Goto-based signing blocks CBMC loop invariants.~{\scriptsize(\href{https://github.com/pq-code-package/mldsa-native/issues/265}{mldsa-native\#265})} \newline SLH-DSA proof uses misdefined security notions.~{\scriptsize(\href{https://github.com/Verified-zkEVM/VCVio/issues/550}{VCVio\#550})} \\
\addlinespace[2pt]
\textbf{Resource and format size limits} (C03), 12 (0.9\%) & Keys, signatures, certificates, or ciphertexts exceed a protocol, format, or interface size limit & Large PQC certs exceed OpenSSL's chain limit.~{\scriptsize(\href{https://github.com/openssl/openssl/issues/28720}{openssl\#28720})} \newline Larger PQC ClientHello stalls at middleboxes.~{\scriptsize(\href{https://github.com/golang/go/issues/80573}{go\#80573})} \\
\addlinespace[2pt]
\rowcolor{gray!7}
\textbf{Scheme-level breaks and algorithm status} (C08), 11 (0.9\%) & A scheme loses security or upstream support, affecting code that depends on it & Remove SIKE, no longer considered secure.~{\scriptsize(\href{https://github.com/cloudflare/circl/issues/358}{circl\#358})} \newline The pqcrypto-mlkem crate is unmaintained.~{\scriptsize(\href{https://github.com/jmlepisto/clatter/issues/29}{clatter\#29})} \\
\addlinespace[2pt]
\bottomrule
\end{tabular}
\end{table*}

\noindent\textbf{Classifier validation.} Gemini 3.8 Flash received the final codebook
with eligibility rules, category definitions, open codes, and tie-breakers.
Schema-constrained outputs recorded eligibility, a problem description, an
open code, primary category and sub-category, an evidence quote, and confidence.
We did not refine the prompt on labelled data.

A held-out sample of 316 initially screen-positive issues excluded discovery
records and was stratified by corpus and year, with at most 15 per repository.
Its size followed Cochran's formula~\cite{cochran1977sampling} for a proportion at 95\% confidence and
a $\pm 5\%$ margin, with finite-population correction for 1{,}755 issues.
Both coders independently applied the frozen codebook, with a required
pre-adjudication primary-category $\kappa \geq 0.80$ on issues both deemed eligible.
The classifier was evaluated once against the adjudicated labels.

The acceptance thresholds were primary-category macro-F1 $\geq 0.80$
and per-category F1 $\geq 0.70$ for categories with at least 30 reference
cases. Model rejections counted as errors on human-confirmed challenges.
The full held-out evaluation included rejection as a separate class.
The 95\% confidence intervals used 2{,}000 issue-level bootstrap resamples.

\noindent\textbf{Counting and analysis.} Category counts included only eligible issues,
using adjudicated labels for discovery and held-out records and classifier
labels for the rest. Recurrence was the number of repositories reporting
each category. Category shares used equal issue weights and equal repository weights.

\begin{figure*}[!t]
\centering
\includegraphics[width=\textwidth]{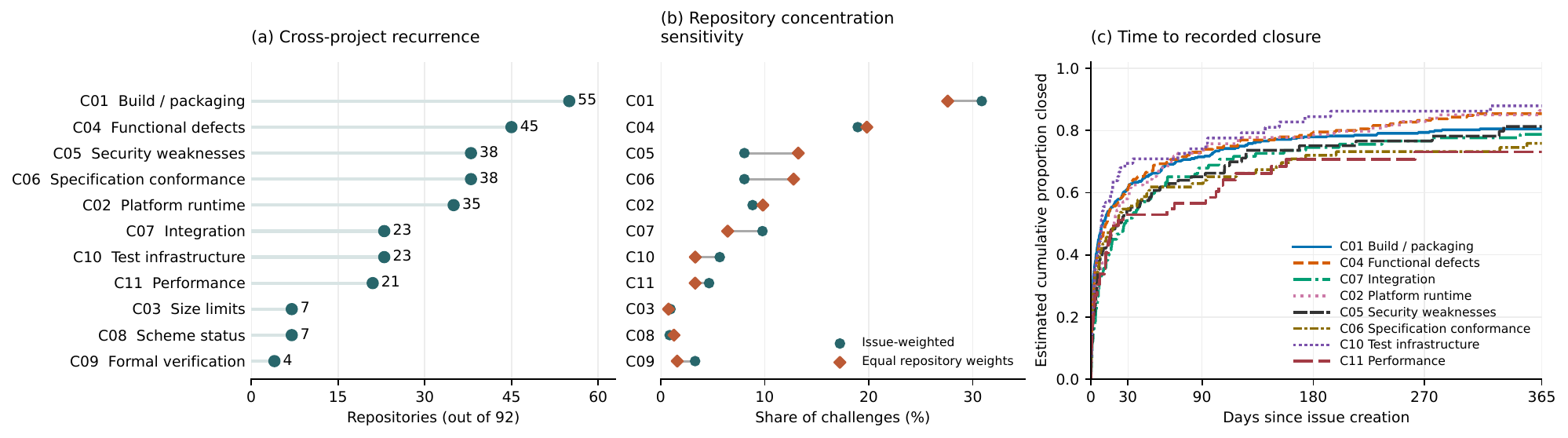}
\vspace{-1.5em}
\caption{Recurrence and closure of PQC technical challenges.
(a) Number of repositories reporting each category.
(b) Category shares when weighting issues or repositories equally.
(c) Kaplan--Meier estimates of the proportion closed by each day for categories
found in at least 10 repositories, with open issues censored at collection.}
\label{fig:rq2-patterns}
\end{figure*}

We used Kaplan--Meier estimates~\cite{kaplan1958nonparametric} for time from
issue creation to recorded closure, right-censoring issues open at collection.
Curves show the estimated proportion closed by time $t$, $1-\hat{S}(t)$,
assuming non-informative censoring
within categories. Recorded closure did not necessarily mean that the problem
was resolved. The analysis did not reconstruct issue-reopening histories.
Figure~\ref{fig:rq2-patterns}c covers the eight categories
present in at least 10 repositories, representing 1{,}202 issues (94.9\%).
The artifact includes all-category curves with pointwise 95\% intervals and
the validation and analysis scripts. Comparisons were descriptive and
unadjusted, without between-category significance tests.

\subsubsection{Result}

Removing 26 bot-authored issues from the 5{,}713 collected issues left
5{,}687 candidates. Humans screened 600 and Gemini screened the remaining
5{,}087. The 1{,}573 model-selected and 182 human-confirmed issues formed
the 1{,}755 screen-positive pool. We used 350 issues for taxonomy discovery,
316 for held-out evaluation, and 1{,}089 for automated classification.

\noindent\textbf{Screening validation.} On the shared pilot, Krippendorff's nominal
$\alpha$ was 0.925 for PQC relevance, 0.918 for a technical challenge, and
0.903 for eligibility. On the 300 screening held-out issues, eligibility
precision was 0.818, recall 0.889, F1 0.852, and Cohen's $\kappa$ 0.794.
Recall fell below the planned 0.90 target. The artifact reports the full
screening evaluation and evidence-quote checks.

\noindent\textbf{Taxonomy.} Round 1 produced 10 categories; round 2 added performance
and efficiency. Rounds 3 and 4 and held-out coding added none, leaving
11 primary categories and 16 sub-categories
(Table~\ref{tab:rq2-taxonomy}).

On issues both coders deemed eligible, pre-adjudication primary-category $\kappa$
was 0.920 in round 2, 0.904 in round 3, 0.891 in round 4, and 0.857 on the
held-out set (95\% interval 0.805--0.904), exceeding the 0.80 requirement.
Eligibility agreement was lower ($\kappa = 0.457$--$0.797$). Round-1 category names
differed between coders. Agreement on issue groupings was 0.54 (adjusted Rand index).

\noindent\textbf{Classifier validation.} On the 235 human-confirmed held-out
challenges, Gemini achieved macro-F1 0.874 (95\% interval 0.796--0.914),
accuracy 0.800, and $\kappa = 0.771$. It met the macro-F1 threshold and
the per-category requirement for C01 (F1 0.847) and C04 (0.833), the only
categories with at least 30 cases.

The model rejected 30 confirmed challenges and accepted 17 of the 81
human-rejected issues (eligibility $\kappa = 0.629$). On the 205 challenges
that both humans and the model deemed eligible, primary-category accuracy was
91.7\%; sub-category accuracy was
90.4\% where the primary category was correct. Across all 316 held-out issues,
including rejection as a class, macro-F1 was 0.841 and $\kappa$ was 0.761.

\noindent\textbf{Category counts.} Gemini categorized 755 of the 1{,}089 remaining
issues, rejected 333, and returned no label for one. Taxonomy classification
cost \$9.55, comprising \$2.07 for held-out evaluation and \$7.48 for the
remaining issues. Combining human and model labels yielded 1{,}267 challenges,
with 1{,}072 closed and 195 open at collection.

\subsubsection{Observation}

Build, packaging, and dependency failures were the largest category,
with 391 issues (30.9\%), followed by incorrect results and functional
defects with 240 (18.9\%). These categories appeared in 55 and 45 of the
92 represented repositories (Figure~\ref{fig:rq2-patterns}a) and remained
the largest under equal repository weights, at 27.6\% and 19.8\%
(Figure~\ref{fig:rq2-patterns}b). Security weaknesses (C05) and specification
conformance and interoperability (C06) each appeared in 38 repositories,
compared with 23 for integration despite its higher issue count.
Formal-verification obstacles were concentrated in four repositories.

Estimated median time to recorded closure was 14.8 days overall,
9.9 days for build failures, and 26.9 days for integration challenges
(Figure~\ref{fig:rq2-patterns}c; observation counts in
Table~\ref{tab:rq2-observation-counts}). Performance had the largest open
share, at 27.1\%. Comparisons did not control for project composition or issue age.

\begin{table}[!t]
\centering
\caption{Issues still open and under observation by day since creation
for Figure~\ref{fig:rq2-patterns}c.}
\label{tab:rq2-observation-counts}
\footnotesize
\setlength{\tabcolsep}{4pt}
\begin{tabular}{@{}lrrrrr@{}}
\toprule
 & \multicolumn{5}{c}{Days since issue creation} \\
\cmidrule(l){2-6}
Category & 0 & 90 & 180 & 270 & 365 \\
\midrule
C01 Build / packaging & 391 & 106 & 81 & 72 & 62 \\
C04 Functional defects & 240 & 55 & 40 & 31 & 25 \\
C07 Integration & 124 & 34 & 26 & 22 & 18 \\
C02 Platform runtime & 112 & 26 & 22 & 16 & 10 \\
C05 Security weaknesses & 102 & 30 & 17 & 15 & 12 \\
C06 Specification conformance & 102 & 33 & 24 & 22 & 18 \\
C10 Test infrastructure & 72 & 15 & 9 & 8 & 7 \\
C11 Performance & 59 & 23 & 13 & 11 & 9 \\
\bottomrule
\end{tabular}
\end{table}

The classifier rejected 30 human-confirmed challenges, including 12 concerning
builds and 7 integration. These categories may therefore be undercounted
among the 1{,}089 automatically classified issues. Nine categories had fewer
than 30 held-out cases and were exempt from the per-category F1 requirement;
their scores are reported with support.

\rqsummary{RQ2}{Build, packaging, and dependency failures (30.9\% of issues)
and functional defects (18.9\%) led the 11-category taxonomy. They recurred
in 55 and 45 repositories and remained largest under equal repository weights.
Integration challenges were less frequent but took longer to reach recorded
closure than build failures (estimated medians of 26.9 versus 9.9 days).}

\subsection{RQ3 Temporal Evolution}
\label{sec:rq3}
\label{sec:results-rq3}
\label{sec:methodology-temporal}

\subsubsection{Approach}

\emph{PQC development activity} comprised RQ1 commits with at least one
\emph{implemented} or \emph{maintenance} label. Commits linked to the same
merged PR counted as one development event; unlinked commits counted separately.
\emph{First adoption} was the earliest \emph{implemented} event for each
repository--scheme pair. The main analysis separated predecessor and standardized
names; a sensitivity analysis combined each lineage at its earliest adoption
date per repository.

Events were dated by PR merge time or commit time for unlinked commits, with
commit and author dates as sensitivity checks. The temporal sample ended in
August 2026, the last complete month before collection. It retained
23{,}513 of RQ1's 24{,}478 adoption-evidence commits, forming 13{,}429 development
events in 134 repositories and 488 first-adoption pairs in 109 repositories.
RQ1's 834 pairs also included 341 with only \emph{maintenance} labels and
5 first adopted after the cutoff.

\begin{figure*}[!t]
\centering
\includegraphics[width=\textwidth]{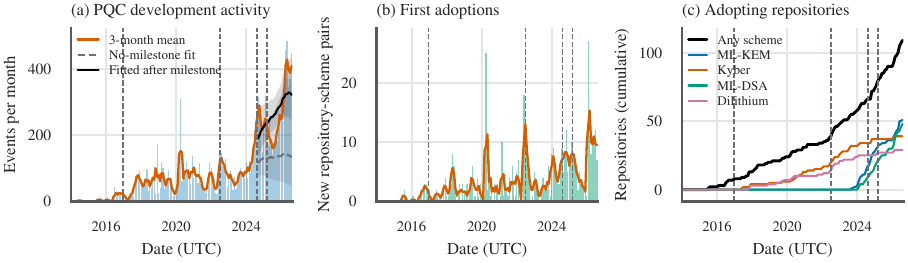}
\vspace{-1.5em}
\caption{Temporal evolution. (a) Monthly adoption-evidence events, with the fit after
the 2024 finalization and its no-milestone counterfactual (95\% bands). (b) New
repository--scheme pairs per month, dated at the first \emph{implemented} event.
(c) Cumulative repositories with a first adoption, overall and for four schemes.
Dashed vertical lines mark the four NIST milestones.}
\label{fig:temporal}
\end{figure*}

The analysis covered the 2016 call, July 2022 selection, August 2024 finalization
of FIPS 203, 204, and 205, and March 2025 HQC selection
(Table~\ref{tab:rq3-models}). Segmented negative-binomial models
\cite{bernal2017its,hilbe2011negativebinomial} described monthly event counts
in eligible milestone windows,
$y_t \sim \mathrm{NB}(\mu_t,\alpha)$,

\begin{equation}
\begin{aligned}
\log \mu_t &= \log e_t + \beta_0 + \beta_1 t\\
&\quad + \beta_2\,\mathrm{post}_t + \beta_3 (t - t_0)\,\mathrm{post}_t,
\end{aligned}
\end{equation}
where $y_t$ is the count in month $t$, $e_t$ is exposure, $t_0$ is the
milestone month, and $\mathrm{post}_t$ indicates subsequent months.
Dispersion $\alpha$ gives $\operatorname{Var}(Y_t)=\mu_t+\alpha\mu_t^2$.
The fitted level ratio at the omitted milestone month is $\exp(\beta_2)$.
The post-milestone monthly growth factor is $\exp(\beta_1+\beta_3)$,
or $\exp(\beta_3)$ times its pre-milestone value. Baseline exposure counted
repositories created by month-end and last pushed at or after month-start.
Temporal dependence in repository data
\cite{saarimaki2022timedependent,robredo2023ignoringtimedependence}
was addressed with Newey--West standard errors~\cite{newey1987hac} and a
three-month bandwidth. Diagnostics compared dispersion
\cite{hilbe2011negativebinomial} and AIC~\cite{akaike1974aic} against Poisson
fits and tested Pearson residuals for autocorrelation using Ljung--Box
tests~\cite{ljung1978lackoffit}.

Windows spanned up to 36 months on each side, excluded the milestone month,
and ended at another primary milestone at least 12 months away.
Each side required at least 12 months and 100 development events or
30 first-adoption events. Milestones within 12 months of a preceding
milestone were excluded.

Intervention starts were set at 0, 3, and 6 months after each milestone
\cite{bernal2018itsmodel}. Shifted-start models assigned intervening months
to the pre-period. Excluded-start models omitted them and retained the
original pre-trend. Autocorrelation sensitivity analyses used Newey--West
bandwidths of 6 and 12 months, generalized estimating equations with AR(1)
working correlation, and a negative-binomial model with a lagged outcome.

Further sensitivity analyses capped each repository at 20 monthly events
and aggregated events quarterly. Alternative exposures counted repositories
with a selected commit that month, all selected commits, and, for first
adoption, repositories without an earlier \emph{implemented} commit.
These analyses assessed sensitivity to GitHub activity biases
\cite{kalliamvakou2016promises}. The estimates describe temporal associations
and do not establish causal effects~\cite{furia2024causal}.

\subsubsection{Result}

Development events increased from 172 in 2016 to 1{,}005 in 2022 and
2{,}915 in the first eight months of 2026, while mean repository exposure
rose from 15 to 118. Monthly activity per repository fell to 0.73 events
in 2023, rose to 1.58 in 2024, and reached 3.08 in 2026
(Figure~\ref{fig:temporal}a). First adoptions numbered 63 in 2022, 31 in
2023, 79 in 2024, 60 in 2025, and 87 in the first eight months of 2026
(Figure~\ref{fig:temporal}b,c).

\begin{table}[!t]
\centering
\caption{Negative-binomial milestone estimates using integration dates.
Incidence rate ratios (95\% intervals) describe immediate level changes
and changes in monthly growth.}
\label{tab:rq3-models}
\footnotesize
\setlength{\tabcolsep}{3pt}
\begin{tabular}{@{}llcc@{}}
\toprule
Milestone & Outcome & Level ratio & Growth ratio \\
\midrule
2016 call (Dec 20) & Development & 1.17 (0.41--3.31) & 0.92 (0.86--0.98) \\
 & First adoption & \multicolumn{2}{c}{not estimable} \\
2022 selection (Jul 5) & Development & 1.17 (0.63--2.18) & 1.03 (0.99--1.07) \\
 & First adoption & 0.69 (0.39--1.22) & 1.01 (0.95--1.08) \\
2024 FIPS (Aug 13) & Development & 1.59 (1.09--2.31) & 1.02 (0.98--1.06) \\
 & First adoption & 1.13 (0.63--2.03) & 1.01 (0.96--1.06) \\
2025 HQC (Mar 11) & \multicolumn{3}{l}{not estimated (within 12 months of 2024 FIPS)} \\
\bottomrule
\end{tabular}
\end{table}

Negative-binomial fits had lower dispersion statistics (1.1--1.6 versus
15--27) and AIC than Poisson fits, with AIC of 528 versus 1{,}258 for 2024.
Only the 2024 fit showed residual autocorrelation
(Ljung--Box $p = 0.002$).

The 2016 analysis had 192 pre-period development events. At lag 0, only
the 2024 development-level interval excluded 1 (Table~\ref{tab:rq3-models}).
At 3- and 6-month lags, the 2022 first-adoption ratios were
0.30 (0.13--0.72) and 0.41 (0.17--0.97), both with intervals below 1.

The 2024 development level ratio was 1.58 with commit dates and 1.59 with
author dates. Repository caps gave 1.38 (1.04--1.84). Active-repository exposure
and quarterly aggregation gave 1.53 (1.15--2.03) and 1.56 (1.10--2.22).
Intervals included 1 under the Poisson model with HAC standard errors
(1.41, 0.90--2.23) and exposure based on all selected commits
(1.17, 0.96--1.41).

The 2024 increase depended on intervention timing and autocorrelation treatment.
Shifting the start by 3 or 6 months
gave level ratios of 0.82 (0.47--1.44) and 0.67 (0.45--1.01). Excluding
those phase-in months gave 1.11 (0.67--1.82) and 0.80 (0.54--1.20).
Newey--West bandwidths of 6 and 12 months retained the 1.59 estimate, with
intervals of 1.11--2.27 and 1.14--2.21. The AR(1) and lagged-outcome
models gave 1.47 (0.86--2.54) and 1.09 (0.80--1.47), with intervals
including 1. The latter showed no residual autocorrelation
(Ljung--Box $p = 0.63$).

The 2024 first-adoption intervals included 1 under commit dates
(1.19, 0.70--2.02), at-risk repository exposure (0.78, 0.39--1.57), and
quarterly aggregation (1.00, 0.48--2.08). Combining predecessor and
standardized names reduced first-adoption pairs from 488 to 436, with similar
estimates at the 2022 selection (0.76, 0.45--1.29) and 2024 finalization
(1.10, 0.63--1.95).

\subsubsection{Observation}

Development events per repository roughly doubled in 2024 and again by 2026.
The milestone estimates depended on model specification. At lag 0,
first-adoption intervals included both no change and increases.

Before the 2022 selection, one of 4{,}406 events used standardized names.
Their share rose to 19\% between selection and finalization and 82\%--84\% afterwards.
Predecessor names accounted for 66\% between selection and finalization,
25\% between finalization and HQC selection, and 11\% thereafter. Events labelled
\emph{hybrid} or \emph{both} accounted for 4\% before the 2022 selection and about 11\%
afterwards. Among repositories with both names, predecessor-first
sequences predominated in every lineage. For Kyber and
ML-KEM, 37 of 50 repositories used Kyber first and 6 used ML-KEM first.
The supplementary material provides full era breakdowns and lineage counts.
Source-level naming patterns did not distinguish algorithm migration
from identifier renaming.

\rqsummary{RQ3}{PQC development activity increased around the 2024 finalization,
but first-adoption rates showed no statistically distinguishable increase
at the estimable milestones. The estimated 2024 increase (ratio 1.59,
95\% interval 1.09--2.31) was sensitive to exposure, timing, and autocorrelation
treatment. Standardized names reached 82\%--84\% of events after finalization,
while predecessor names persisted.}

\section{Discussion}
\label{sec:discussion}

Maintenance accounted for 90.6\% of scheme-level adoption-status labels and new
implementations for 3.8\%. Build failures and functional defects remained the
largest challenge categories when repositories received equal weight
(Figure~\ref{fig:rq2-patterns}a--b). \textbf{Maintainers and adopters should budget for maintenance beyond
initial implementation, including build compatibility and
functional correctness as schemes and dependencies change.}

Security weaknesses and specification-conformance issues each appeared in 38
repositories, while the more numerous integration issues appeared in 23
(Figure~\ref{fig:rq2-patterns}a). Issue frequency and cross-project recurrence
therefore rank the categories differently. \textbf{Researchers should evaluate PQC development tools by
challenge category and report both issue-weighted and repository-weighted
performance.}

\textbf{Cryptographic inventories should record scheme names, lineages,
deployment modes, classical partners, and observation dates.} In each renamed
lineage, at least half of repositories with standardized scheme names also had
predecessor names in their commit histories
(Figure~\ref{fig:scheme-name-coexistence}). Predecessors and standardized successors should remain
separate entries linked by lineage. Historical coexistence does not establish
simultaneous support, which requires evidence tied to the same release or revision.

Development events include maintenance, while first adoption counts only the
earliest observed implementation per repository--scheme pair. Neither
establishes deployment (Figures~\ref{fig:corpus-scheme-patterns} and~\ref{fig:temporal}). Around the
2024 finalization, development activity rose while first adoption showed no
statistically distinguishable change. Reporting activity alone risks conflating
maintenance with first adoption. \textbf{Progress reports should
distinguish PQC development activity, first adoption, and deployment.}
The estimated increase depends on model specification and does not establish causation
(Section~\ref{sec:rq3}).

\section{Related Work}
\label{sec:related-work}

Binary analysis of more than 4{,}000 Android apps found no adoption of
standardized PQC \cite{strauss2025assessingenhancingquantumreadiness},
whereas TLS measurements found deployment concentrated on hybrid
X25519MLKEM768 and shaped by major infrastructure providers
\cite{loizou2026measuringpostquantumtlsdeployment,
wickramasinghe2026mindgappolicyvs}. Static analysis identified
PQC API usage in Python, Java, and Go repositories
\cite{mehl2026pqcreadiness}. These measures capture different populations and
stages of adoption.

Migration studies assess cryptographic dependencies and inventory assets
\cite{costa2026cryptoagilityreadiness,rameshan2026assessmentframework,
bogli2026bfcbom}. Reviews organize the transition process
\cite{nather2024migratingsoftwaresystems}, while toolchains and LLM-assisted
code-fragment migration support implementation
\cite{rattanavipanon2025toolchain,debonrostro2026empiricalevaluationlargelanguage}.
Android benchmarks assess implementations on target devices
\cite{chatterjee2026benchmarkingintegrationandroid}. Commit and issue histories
complement these assessments with evidence of how projects implement and
maintain PQC support. Our study uses these histories to identify recurring
technical challenges and trace adoption around NIST milestones.

\section{Threats to Validity}
\label{sec:threats}

\emph{Construct validity.} Keyword retrieval can miss PQC work or include
unrelated changes despite multiple discovery channels and relevance assessment
from patches and threads. Adoption labels describe source-level implementation
and maintenance, not deployed software; issue closure does not establish resolution.

\emph{Internal validity and reliability.} Classification remains subject
to human and model error. Agreement checks covered three RQ1 pilot annotators
and two RQ2 taxonomy coders; a third researcher adjudicated RQ2 disagreements.
Codebooks were frozen before held-out evaluation. The artifact preserves
prompts, labels, raw outputs, and failures.

\emph{Conclusion validity.} Rare-category estimates are less stable;
reported class counts and intervals document this uncertainty. Model comparisons
use multiplicity corrections. Unweighted proportions from enriched validation samples do not
estimate corpus prevalence.
Temporal analyses adjust for repository counts and vary exposure, lags, and
autocorrelation treatment. Milestone associations do not establish causality.

\emph{External validity.} Findings apply to the sampled public GitHub
repositories; private projects, other platforms, and undiscovered repositories
may differ. Corpus~B's keyword-matched artifacts do not cover all PQC activity
in general-purpose software.

\section{Conclusion}
\label{sec:conclusion}

This study examined PQC adoption and recurring technical challenges through
commit histories and issue threads from 179 GitHub repositories. Manual coding
established an 11-category taxonomy; a validated LLM classified the remaining
commits and issues.
Maintenance dominated scheme-level changes, while build failures and functional
defects recurred across projects. Standardized and predecessor names coexisted, and
increased PQC development activity around the 2024 finalization was not
accompanied by a statistically distinguishable increase in first-adoption rates.
These findings support migration plans that account for continued maintenance,
build reliability, and compatibility testing beyond initial implementation.

\section*{Data Availability Statement}
\label{sec:data-availability}
We release an anonymized replication package~\cite{pqcAdoptionReplicationPackage}. It contains the analysis
pipeline, records, prompts, model outputs, codebooks, and annotations behind both analyses.

\balance
\bibliographystyle{IEEEtran}
\bibliography{references}

\end{document}